\documentclass[12pt]{article}
\usepackage{amsmath, amsgen,amstext,amsbsy,amsopn,amsthm,amssymb,graphicx}

\theoremstyle{definition}

\newcommand{\R}{\mathbb{R}}

\newcommand{\const}{{\rm const. \,}}

\def\R{{\mathbb R}}

\def\x{{\bf  x}}
\def\p{{\bf p}}

\def\y{{\bf y}}

\def\0{{\bf 0}}

\def\const{{\rm const.\,}}

\def\mfr#1/#2{\hbox{$\frac{{#1}}{{#2}}$}}
\def\uprho{\raise1pt\hbox{$\rho$}}
\def\upchi{\raise1pt\hbox{$\chi$}}
\def\dlambda{\lower1pt\hbox{$\lambda$}}

\newcommand{\Z}{\mathbb{Z}}

\newcommand{\beq}{\begin{equation}}
\newcommand{\eeq}{\end{equation}}

\newcommand{\eps}{\varepsilon}

\newcommand{\Tr}{{\rm Tr}\, }
\newcommand{\half}{\mbox{$\frac{1}{2}$}}

\newcommand{\vecp}{{\p}}

\newcommand{\vecx}{{\x}}
\newcommand{\vecy}{{\y}}

\begin{document}
\title{\bf{Bose-Einstein Condensation \\ as a Quantum Phase 
Transition \\ in an Optical Lattice$^*$}}
\author{\vspace{5pt} M. Aizenman$^1$, E.H.~Lieb$^{1}$, R. 
Seiringer$^{1}$, J.P. Solovej$^2$ and J.
Yngvason$^{3}$\\
\vspace{-4pt}\small{$1.$ Departments of Mathematics and Physics,
Jadwin Hall,} \\
\small{Princeton University, P.~O.~Box 708, Princeton, New Jersey
   08544}\\
\vspace{-4pt}\small{$2.$ Department of Mathematics,
University of Copenhagen,} \\ \small{ Universitetsparken 5, DK-2100 Copenhagen,
Denmark} \\
\vspace{-4pt}\small{$3.$ Institut f\"ur Theoretische Physik, Universit\"at
Wien,}\\
\small{Boltzmanngasse 5, A-1090 Vienna, Austria, and} \\ 
\vspace{-4pt}\small{Erwin Schr\"odinger Institute for Mathematical 
Physics,} \\ \small{Boltzmanngasse 9, A-1090 Vienna, Austria}
}
\date{\small Nov. 11, 2004}
\maketitle

\renewcommand{\thefootnote}{$*$}
\footnotetext{\scriptsize{Contribution to the proceedings of QMath9, 
Giens, France, Sept. 12--16, 2004. Talk given  by Jakob Yngvason.
\copyright\, 2004 by the authors. This paper may be reproduced, in its
entirety, for non-commercial purposes. \\ \vskip -.2cm
Work supported in part by US NSF grants PHY 9971149 (MA), PHY 
0139984-A01 (EHL),
PHY 0353181 (RS) and DMS-0111298 (JPS); by an A.P.~Sloan Fellowship
(RS); by EU grant HPRN-CT-2002-00277 (JPS and JY); by FWF grant
P17176-N02 (JY); by MaPhySto -- A Network in Mathematical Physics and
Stochastics funded by The Danish National Research Foundation (JPS),
and by grants from the Danish research council (JPS).}}

\begin{abstract}
   One of the most remarkable recent developments in the study of
   ultracold Bose gases is the observation of a reversible transition
   from a Bose Einstein condensate to a state composed of localized
   atoms as the strength of a periodic, optical trapping potential is
   varied. In \cite{ALSSY} a model of this phenomenon has been analyzed
   rigorously. The gas is a hard core lattice gas and the optical
   lattice is modeled by a periodic potential of strength $\lambda$.
   For small $\lambda$ and temperature Bose-Einstein condensation (BEC)
   is proved to occur, while at large $\lambda$ BEC disappears, even in
   the ground state, which is a Mott-insulator state with a
   characteristic gap. The inter-particle interaction is essential for
   this effect. This contribution gives a pedagogical survey of these
   results.
\end{abstract}
\bigskip

\section{Introduction}

One of the most remarkable recent developments in the study of
ultracold Bose gases is the observation of a reversible transition
from a Bose-Einstein condensate to a state composed of localized atoms
as the strength of a periodic, optical trapping potential is varied
\cite{G1,G2}.  This is an example of a quantum phase transition
\cite{Sa} where quantum fluctuations and correlations rather than
energy-entropy competition is the driving force and its theoretical
understanding is quite challenging.  The model usually considered for
describing this phenomenon is the Bose-Hubbard model and the
transition is interpreted as a transition between a superfluid and a
{\it Mott insulator} that was studied in \cite{FWGF} with an
application to ${\rm He}^4$ in porous media in mind.  The possibility
of applying this scheme to gases of alkali atoms in optical traps was
first realized in \cite{JBCGZ}.  The article \cite{Z} reviews these
developments and many recent papers, e.g.,
\cite{GCZKSD,Ziegler,NS,G,Zi,DODS,RBREWC,MA,Auer} are devoted to this
topic.  These papers contain also further references to earlier work
along these lines.

The investigations of the phase transition in the Bose-Hubbard model
are mostly based on variational or numerical methods and the signal of
the phase transition is usually taken to be that an ansatz with a
sharp particle number at each lattice site leads to a lower energy
than a delocalized Bogoliubov state.  On the other hand, there exists
no rigorous proof, so far, that the true ground state of the model has
off-diagonal long range order at one end of the parameter regime that
disappears at the other end.  In this contribution, which is based on the
paper \cite{ALSSY}, we study a slightly different model where just
this phenomenon can be rigorously proved and which, at the same time,
captures the salient features of the experimental situation.

Physically, we are dealing with a trapped Bose gas with short range
interaction. The model we discuss, however, is not a continuum model
but rather a lattice gas, i.e., the particles are confined to move on
a $d$-dimensional, hypercubic lattice and the kinetic energy is given
by the discrete Laplacian. Moreover, when discusssing BEC, it is
convenient not to fix the particle number but to work in a
grand-canonical ensemble. The chemical potential is fixed in such a
way that the average particle number equals half the number of lattice
sites, i.e., we consider {\it half filling}.  (This restriction is
dictated by our method of proof.)  The optical lattice is modeled by a
periodic, one-body potential. In experiments the gas is enclosed in an
additional trap potential that is slowly varying on the scale of the
optical lattice but we neglect here the inhomogeneity due to such a
potential and consider instead the thermodynamic limit.

In terms of bosonic creation and annihilation operators,
$a^\dagger_\x$ and $a^{\phantom\dagger}_\x$, our Hamiltonian is
expressed as
\begin{equation}
H= - \half \sum_{\langle \vecx\vecy\rangle} ( a^\dagger_\vecx
a^{\phantom\dagger}_\vecy + a^{\phantom\dagger}_\vecx a^\dagger_\vecy
) + \lambda \sum_\vecx (-1)^\vecx a^\dagger_\vecx a^{\phantom\dagger}_\vecx
+ U \sum_\vecx a^\dagger_\vecx a^{\phantom\dagger}_\vecx
(a^\dagger_\vecx a^{\phantom\dagger}_\vecx-1). \label{hamiltonian}
\end{equation}
The sites $\vecx$ are in a cube $\Lambda\subset \Z^d$ with opposite
sides identified (i.e., a $d$-dimensional torus) and $\langle
\vecx\vecy\rangle$ stands for pairs of nearest neighbors. Units are 
chosen such that $\hbar^2/m =1$.

The first term in \eqref{hamiltonian} is the discrete Laplacian
$\sum_{\langle
\vecx\vecy\rangle}(a^\dagger_\x-a^\dagger_\y)
(a^{\phantom\dagger}_\x-a^{\phantom\dagger}_\y)$
minus $2d\sum_{\x} a^\dagger_\vecx a^{\phantom\dagger}_\x$, i.e., we
have subtracted a chemical potential that equals $d$.

The optical lattice gives rise to a potential $\lambda (-1)^{\x}$
which alternates in sign between the $A$ and $B$ sublattices of even
and odd sites.  The inter-atomic on-site repulsion is $U$, but we consider
here only the case of a {\it hard-core interaction}, i.e., $U=\infty$. If
$\lambda = 0$ but $U < \infty$ we have the Bose-Hubbard model.  Then
all sites are equivalent and the lattice represents the attractive
sites of the optical lattice. In our case the adjustable parameter is
$\lambda$ instead of $U$ and for large $\lambda$ the atoms will try to
localize on the $B$ sublattice. The Hamiltonian \eqref{hamiltonian}
conserves the particle number $N$ and it can be shown that, for 
$U=\infty$, the lowest
energy is obtained uniquely for $N=\half|\Lambda|$, i.e., half the number of
lattice sites.  Because of the periodic potential the unit cell in
this model consists of two lattice sites, so that we have on average
one particle per unit cell. This corresponds, physically, to filling
factor 1 in the Bose-Hubbard model.

For given temperature $T$, we consider grand-canonical thermal
equilibrium states, described by the Gibbs density matrices
$Z^{-1}\exp(-\beta H)$ with $Z$ the normalization factor (partition
function) and $\beta=1/T$ the inverse temperature. Units are chosen so
that Boltzmann's constant equals 1. The thermal expectation value of
some observable $\mathcal{O}$ will be denoted by $\langle{\mathcal
   O}\rangle=Z^{-1}\Tr {\mathcal O}\exp(-\beta H)$.

Our main results about this model can be summarized as follows:

\begin{itemize}

\item[1.]\label{item1}
If $T$ and $\lambda$ are both small, there is Bose-Einstein
condensation.  In this parameter regime the one-body density matrix
$\gamma(\x,\y)=\langle a^\dagger_\x a^{\phantom
\dagger}_\y
\rangle$
has exactly one large eigenvalue (in the thermodynamic limit), and the
corresponding condensate wave function is $\phi(\x)=$constant.

\item[2.]\label{item2} If either $T$ or $\lambda$ is big enough, then
   the one-body density matrix decays exponentially with the distance
   $|\x-\y|$, and hence there is {\it no BEC}.  In particular, this
   applies to the ground state $T=0$ for $\lambda$ big enough, where
   the system is in a Mott insulator phase.

\item[3.]\label{item3}
The Mott insulator phase is characterized by a gap, i.e., a jump in
the chemical potential. We are able to prove this, at half-filling, in
the region described in item~2 above.  More precisely, there
is a cusp in the dependence of the ground state energy on the number
of particles; adding or removing one particle costs a non-zero amount
of energy.  We also show that there is no such gap whenever there is
BEC.

\item[4.] The interparticle interaction is essential for items~2 and~3.
Non-interacting bosons {\it always display BEC} for low, but
   positive $T$ (depending on $\lambda$, of course).

\item[5.] For all $T\geq 0$ and all $\lambda > 0$ the diagonal part of
   the one-body density matrix $\langle a^\dagger_\x a^{\phantom
     \dagger}_\x \rangle$ (the one-particle density) is {\it not
     constant}. Its value on the A sublattice is constant, but strictly
   less than its constant value on the B sublattice and this
   discrepancy survives in the thermodynamic limit. In contrast, in the
   regime mentioned in item~1, the off-diagonal long-range
   order is constant, i.e., $\langle a^\dagger_\x a^{\phantom
     \dagger}_\y \rangle \approx \phi(\x) \phi(\y)^*$ for large
   $|\x-\y|$ with $\phi(\x)=$constant.
\end{itemize}

\begin{figure}[htf]
\center
\includegraphics[width=9cm, height=6.6cm]{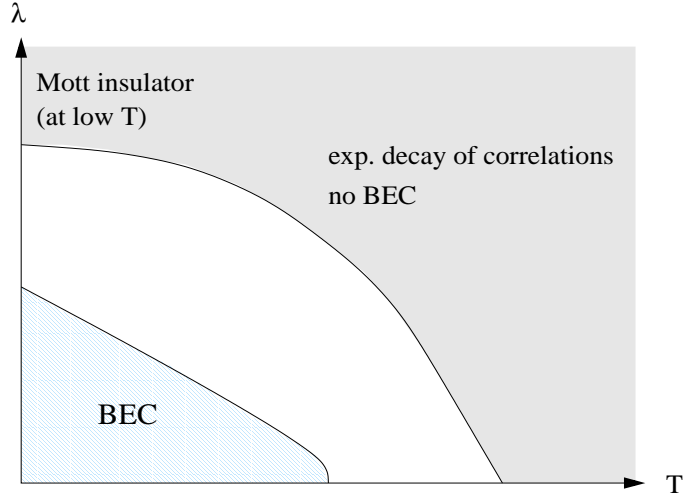}
\caption{Schematic phase diagram at half-filling}
\label{fig1}
\end{figure}

Because of the hard-core interaction between the particles, there is
at most one particle at each site and our Hamiltonian (with
$U=\infty$) thus acts on the Hilbert space ${\mathcal
H}=\bigotimes_{\x\in\Lambda}{\mathbb C}^2$.  The creation and
annihilation operators can be represented as $2\times 2$ matrices with
$$ a^\dagger_{\vecx}\leftrightarrow \left( \begin{array}{cc} 0 & 1 \\
0 & 0
\end{array} \right) , \quad a^{\phantom\dagger}_{\vecx}\leftrightarrow
\left( \begin{array}{cc} 0 & 0 \\ 1 & 0 \end{array} \right) , \quad
a^\dagger_{\vecx}a^{\phantom\dagger}_\vecx\leftrightarrow \left(
\begin{array}{cc} 1 & 0 \\ 0 & 0 \end{array} \right),
$$
for each $\vecx\in\Lambda$. More precisely, these matrices act on
the tensor factor associated with the site $\x$ while
$a^\dagger_{\vecx}$ and $a^{\phantom\dagger}_{\vecx}$ act as the 
identity on the other
factors in the Hilbert space ${\mathcal
H}=\bigotimes_{\x\in\Lambda}{\mathbb C}^2$.

The Hamiltonian can alternatively be written in terms of the spin 1/2
operators $$ S^1=\frac12 \left(
\begin{array}{cc} 0 & 1 \\ 1 & 0 \end{array} \right), \quad S^2=\frac12\left(
\begin{array}{cc} 0 & -{\rm i} \\ {\rm i} & 0 \end{array} \right), \quad
S^3=\frac12\left( \begin{array}{cc} 1 & 0 \\ 0 & -1 \end{array}
\right). $$ The correspondence with the creation and annihilation
operators is $$ a^\dagger_{\vecx}=S^1_{\vecx}+{\rm i}S^2_{\vecx}\equiv
S_\vecx^+, \quad a^{\phantom\dagger}_{\vecx}=S^1_{\vecx}-{\rm
i}S^2_{\vecx}\equiv S_\vecx^-, $$ and hence
$a^\dagger_{\vecx}a^{\phantom\dagger}_{\vecx}=S_\vecx^3+\half$.  (This
is known as the Matsubara-Matsuda correspondence \cite{MM}.)  Adding a
convenient constant to make the periodic potential positive, the
Hamiltonian (\ref{hamiltonian}) for $U=\infty$ is thus equivalent to
\begin{eqnarray}\label{hamspin}
H&=&-\half \sum_{\langle \vecx\vecy\rangle} (S^+_\vecx
S^-_\vecy+S^-_\vecx S^+_\vecy)+\lambda\sum_\vecx \left[\half +
(-1)^\vecx S^3_\vecx\right]
\nonumber\\
&=&-\sum_{\langle \vecx\vecy\rangle}(S^1_\vecx S^1_\vecy+S^2_\vecx
S^2_\vecy) +\lambda\sum_\vecx\left[ \half + (-1)^\vecx
S^3_\vecx\right].
\end{eqnarray}
Without loss of generality we may assume $\lambda\geq 0$. This
Hamiltonian is well known as a model for interacting spins, referred
to as the XY model~\cite{DLS}. The last term has the interpretation of a
staggered magnetic field. We note that BEC for the lattice gas is
equivalent to off-diagonal long range order for the 1- and 2-components
of the spins.

The Hamiltonian \eqref{hamspin} is clearly invariant under
simultaneous rotations of all the spins around the 3-axis. In particle
language this is the $U(1)$ gauge symmetry associated with particle
number conservation of the Hamiltonian \eqref{hamiltonian}.
Off-diagonal long range order (or, equivalently, BEC) implies that
this symmetry is spontaneously broken in the state under
consideration. It is notoriously difficult to prove such symmetry
breaking for systems with a continuous symmetry. One of the few
available techniques is that of {\it reflection positivity} (and the
closely related property of {\it Gaussian domination}) and fortunately
it can be applied to our system. For this, however, the hard core and
half-filling conditions are essential because they imply a
particle-hole symmetry that is crucial for the proofs to work.
Naturally, BEC is expected to occur at other fillings, but no one has
so far found a way to prove condensation (or, equivalently, long-range
order in an antiferromganet with continuous symmetry) without using
reflection positivity and infrared bounds, and these require the
addtional symmetry.

Reflection positivity was first formulated by K.\ Osterwalder and R.\
Schrader \cite{oster} in the context of relativistic quantum field theory.
Later, J. Fr\"ohlich, B. Simon and T. Spencer used the concept to
prove the existence of a phase transition for a classical spin model
with a continuous symmetry \cite{FSS}, and E.\ Lieb and J.\ Fr\"ohlich
\cite{FL} as well as F.\ Dyson, E.\ Lieb and B.\ Simon \cite{DLS}
applied it for the analysis of quantum spin systems. The proof of
off-diagonal long range order for the Hamiltonian \eqref{hamspin} (for
small $\lambda$) given here is based on appropriate modifications of
the arguments in \cite{DLS}.

\section{Reflection Positivity}

In the present context reflection positivity means the following. We
divide the torus $\Lambda$ into two congruent parts, $\Lambda_{\rm L}$
and $\Lambda_{\rm R}$, by cutting it with a hyperplane orthogonal to
one of the $d$ directions. (For this we assume that the side length of
$\Lambda$ is even.) This induces a factorization of the Hilbert space,
${\mathcal H}={\mathcal H}_{\rm L}\otimes {\mathcal H}_{\rm R}$, with
$${\mathcal H}_{\rm L,R}=\bigotimes_{\x\in\Lambda_{\rm L,R}}{\mathbb
C}^2.$$ There is a natural identification between a site
$\x\in\Lambda_{\rm L}$ and its mirror image $\vartheta\x
\in\Lambda_{\rm R}$. If $F$ is an operator on ${\mathcal H}={\mathcal
   H}_{\rm L}$ we define its reflection $\theta F$ as an operator on
   ${\mathcal H}_{\rm R}$ in the following way. If $F=F_\x$ operates
   non-trivially only on one site, $\x\in \Lambda_{\rm L}$, we define
   $\theta F=VF_{\vartheta\x}V^\dagger$ where $V$ denotes the unitary
   particle-hole transformation or, in the spin language, rotation by
   $\pi$ around the 1-axis. This definition extends in an obvious
   way to products of operators on single sites and then, by linearity,
   to arbitrary operators on ${\mathcal H}_{\rm L}$. Reflection
   positivity of a state $\langle\,\cdot\,\rangle$ means that
\begin{equation}\label{reflpos}
\langle
  F\theta \overline F\rangle \geq 0
\end{equation}
  for any $F$ operating on
   ${\mathcal H}_{\rm L}$. Here $\overline F$ is the complex
   conjugate of the operator $F$ in the matrix representation defined
   above, i.e., defined by the basis where the operators $S^3_{\x}$ are
   diagonal.

We now show that reflection positivity holds for any thermal
equilibrium state of our Hamiltonian. We can write the
Hamiltonian~(\ref{hamspin}) as
\beq\label{hamreal}
H=H_{\rm L} + H_{\rm R}-\half \sum_{\langle \vecx\vecy\rangle\in 
M}(S^+_\vecx S^-_\vecy +S^-_\vecx
S^+_\vecy),
\eeq
where $H_{\rm L}$ and $H_{\rm R}$ act non-trivially
only on ${\mathcal H}_{\rm L}$ and ${\mathcal H}_{\rm R}$,
respectively. Here, $M$ denotes the set of bonds going from the left
sublattice to the right sublattice. (Because of the periodic
boundary condition these include the bonds that connect the right
boundary with the left boundary.) Note that $H_{\rm R} = \theta H_{\rm L}$, and
$$
\sum_{\langle \vecx\vecy\rangle\in M}(S^+_\vecx S^-_\vecy +S^-_\vecx
S^+_\vecy)= \sum_{\langle \vecx\vecy\rangle\in M}(S^+_\vecx \theta 
S^+_\vecx +S^-_\vecx
\theta S^-_\vecx).
$$
For these properties it is essential that we included the unitary 
particle-hole transformation $V$ in the definition of the reflection 
$\theta$. For reflection positivity it is also important that  all 
operators appearing in $H$ (\ref{hamreal}) have a {\it real} matrix 
representation. Moreover, the minus sign in (\ref{hamreal}) is 
essential.

Using the Trotter product formula, we have
$$
\Tr F \theta \overline F e^{-\beta H} = \\ \lim_{n\to\infty} \Tr F 
\theta \overline F \, {\mathcal Z}_n
$$
with
\beq\label{trotter}
{\mathcal Z}_n = \left[ e^{-\frac 1n \beta H_L} \theta e^{-\frac 1n 
\beta H_L} \prod_{\langle \vecx\vecy\rangle\in 
M}\left(1+\frac{\beta}{2n} \left[ S^+_\vecx \theta S^+_\vecx 
+S^-_\vecx
\theta S^-_\vecx)\right] \right) \right]^n.
\eeq
Observe that ${\mathcal Z}_n$ is a sum of terms of the form
\beq\label{aaa}
\mbox{$\prod_i$} A_i \theta A_i,
\eeq
with $A_i$ given by either $e^{-\frac 1n \beta H_L}$ or 
$\sqrt{\frac{\beta}{2n}}S^+_\vecx$ or 
$\sqrt{\frac{\beta}{2n}}S^-_\vecx$. All the $A_i$ are real matrices, 
and therefore
\beq
{\rm Tr}_{\mathcal H}\,  F \theta \overline F\, \mbox{$\prod_i $}A_i 
\theta  A_i =
{\rm Tr}_{\mathcal H}\,  F \mbox{$\prod_i $}A_i \, \theta\left[ 
\overline F \mbox{$\prod_j$} A_j\right] = \big| {\rm Tr}_{{\mathcal 
H}_{\rm L}}\, F \mbox{$\prod_i$} A_i \big|^2 \geq 0.
\eeq
Hence $\Tr F\theta\overline F\, {\mathcal Z}_n$  is a sum of 
non-negative terms and therefore non-negative. This proves our 
assertion.

\section{Proof of BEC for Small $\lambda$ and $T$}

The main tool in our proof of BEC are {\it infrared bounds}. More 
precisely, for $\p\in \Lambda^*$ (the dual lattice of $\Lambda$), let 
$\widetilde S^\#_\p=|\Lambda|^{-1/2} \sum_\x S_\x^\# \exp({\rm i} 
\p\cdot
\x)$ denote the Fourier transform of the spin operators. We claim that
\beq\label{infb}
(\widetilde S_\p^1, \widetilde S^1_{-\p}) \leq \frac{T}{ 2 E_\vecp},
\eeq
with $E_\vecp= \sum_{i=1}^d (1-\cos(p_i))$. Here, $p_i$ denotes the 
components of $\vecp$, and
$(\, ,\, )$ denotes the Duhamel two point function at temperature $T$,
defined by
\beq
(A,B)=\int _0^1 \Tr\left( A e^{-s \beta H} B e^{-(1-s)\beta H}
\right) ds / \Tr e^{-\beta H}
\eeq
for any pair of operators $A$ and $B$.
Because of invariance under rotations around the $S^3$ axis, 
(\ref{infb}) is equally true with $S^1$ replaced by $S^2$, of course.

The crucial lemma ({\it Gaussian domination}) is the following.
Define, for a complex valued function $h$ on the bonds $\langle 
\vecx\vecy \rangle$ in $\Lambda$,
\beq
Z(h)=\Tr \exp\left[ - \beta K(h) \right],
\eeq
with $K(h)$ the modified Hamiltonian
\beq
K(h)= \frac 14 \sum_{\langle \vecx\vecy\rangle} \left( 
\big(S^+_\vecx-S^-_\vecy-h_{\vecx\vecy}\big)^2
+ \big(S^-_\vecx-S^+_\vecy-\overline{h_{\vecx\vecy}}\big)^2\right)+ 
\lambda\sum_\vecx \big[\half + (-1)^\vecx S^3_\vecx\big].
\eeq
Note that for $h\equiv 0$, $K(h)$ agrees with the Hamiltonian $H$, 
because $(S^{\pm})^2=0$. We claim that, for any {\em real valued} $h$,
\beq\label{gauss}
Z(h)\leq Z(0).
\eeq
The infrared bound then follows from $d^2 Z(\varepsilon
h)/d\varepsilon^2|_{\eps=0}\leq 0$, taking $h_{\x\y}=\exp({\rm 
i}\p\cdot \x)- \exp({\rm i}\p\cdot \y)$. This is not a real function, 
though, but the negativity of the (real!) quadratic form $d^2 
Z(\varepsilon
h)/d\varepsilon^2|_{\eps=0}$ for real $h$ implies negativity also for 
complex-valued $h$.

The proof of (\ref{gauss}) is very similar to the proof of the
reflection positivity property (\ref{reflpos}) given above. It follows
along the same lines as in \cite{DLS}, but we repeat it here for
convenience of the reader.

The intuition behind (\ref{gauss}) is the following. First, in
maximizing $Z(h)$ one can restrict to gradients, i.e., $h_{\x\y}= \hat
h_\x-\hat h_\y$ for some function $\hat h_\x$ on $\Lambda$. (This
follows from stationarity of $Z( h)$ at a maximizer $h_{\rm max})$.)
Reflection positivity implies that $\langle A\theta \overline
B\rangle$ defines a scalar product on operators on $\mathcal H_{\rm
   L}$, and hence there is a corresponding Schwarz inequality.
Moreover, since reflection positivity holds for reflections across
{\it any} hyperplane, one arrives at the so-called {\it chessboard
   inequality}, which is simply a version of Schwarz's inequality for
multiple reflections across different hyperplanes. Such a chessboard
estimate implies that in order to maximize $Z(h)$ it is best to choose
the function $\hat h_\vecx$ to be constant. In the case of classical
spin systems \cite{FSS}, this intuition can be turned into a complete
proof of (\ref{gauss}).
Because of non-commutativity of $K(h)$
with $K(0)=H$, this is not possible in the quantum case. However, one
can proceed by using the Trotter formula as follows.

Let $h_{\rm max}$ be a function that maximizes $Z(h)$ for real valued 
$h$. If there is more than one maximizer, we choose $h_{\rm max}$ to 
be one that vanishes on the largest number of bonds. We then have to 
show that actually $h_{\rm max}\equiv 0$. If $h_{\rm max}\not\equiv 
0$, we draw a hyperplane such that $h_{\vecx\vecy}\neq 0$ for at 
least one pair $\langle \vecx\vecy\rangle$ crossing the plane. We can 
again write
\beq
K(h) = K_L(h) + K_R(h)  + \frac 14 \sum_{\langle \vecx\vecy\rangle\in 
M} \left(  (S^+_\vecx-S^-_\vecy-h_{\vecx\vecy})^2
+ (S^-_\vecx-S^+_\vecy-h_{\vecx\vecy})^2\right).
\eeq
Using the Trotter formula, we have  $Z(h)=\lim_{n\to\infty} \alpha_n$, 
with
\beq\label{tro}
\alpha_n= \Tr \left[ e^{-\beta K_L/n} e^{-\beta K_R/n} \prod_{\langle 
\x\y\rangle\in M} e^{-\beta (S^+_\vecx-S^-_\vecy-h_{\vecx\vecy})^2 /4 
n} e^{-\beta (S^-_\vecx-S^+_\vecy-h_{\vecx\vecy})^2 /4 n}\right]^n .
\eeq
For any matrix, we can write
\beq
e^{-D^2} = (4\pi)^{-1/2} \int_\R dk\, e^{{\rm i}k D} e^{-k^2/4}.
\eeq
If we apply this to the last two factors in (\ref{tro}), and note 
that $S^-_\y = \theta S^+_\x$ if $\langle\x\y\rangle\in M$. Denoting 
by $\x_1, \dots, \x_l$ the points on the left side of the bonds in 
$M$, we have that
\begin{eqnarray}\nonumber
\alpha_n &=& (4\pi)^{-n l} \int_{R^{2nl}} d^{2nl} k \, \Tr
\left[  e^{-\beta K_L/n} e^{-\beta K_R/n} e^{{\rm i} k_1 (S^+_{\x_1} 
- \theta S^+_{\x_1}) \beta^{1/2} /2  n^{1/2}}  \dots \right] \\ && 
\times e^{- k^2/4 } e^{-{\rm i} k_1 h_{\x_1 \vartheta \x_1} 
\beta^{1/2}/2n^{1/2} \dots}.
\end{eqnarray}
Here we denotes $k^2 =\sum k_i^2$ for short.
Since matrices on the right of $M$ commute with matrices on the left, 
and since all matrices in question are {\it real}, we see that the 
trace in the integrand above can be written as
\beq
\Tr \left[  e^{-\beta K_L/n} e^{{\rm i} k_1 S^+_{\x_1} \beta^{1/2}/2 
n^{1/2}}  \dots \right] \overline{ \Tr \left[  e^{-\beta K_R/n} 
e^{{\rm i} k_1 \theta S^+_{\x_1} \beta^{1/2} /2 n^{1/2}}  \dots 
\right]}.
\eeq
Using the Schwarz inequality for the $k$ integration, and \lq 
undoing\rq\ the above step, we see that
\begin{eqnarray}\nonumber
\!\!\!\!\! |\alpha_n|^2 \!\!\! &\leq& \!\!\!\left( (4\pi)^{-n l} 
\int_{R^{2nl}} d^{2nl} k \, e^{- k^2/4} \right.\\ \nonumber && \left. 
\quad \times\Tr
\left[  e^{-\beta K_L/n} e^{-\beta \theta K_L/n} e^{{\rm i} k_1 
(S^+_{\x_1} - \theta S^+_{\x_1})\beta^{1/2} /2 n^{1/2}}  \dots 
\right] \phantom{\int_{R^{2nl}} } \!\!\!\!\! \!\!\!\!\!\!\!\!\! 
\right) \\ \nonumber && \!\!\!\!\!\times  \left( (4\pi)^{-n l} 
\int_{R^{2nl}} d^{2nl} k \, e^{- k^2/4} \right.\\ && \left. \quad 
\times \Tr
\left[  e^{-\beta \theta K_R/n} e^{-\beta K_R/n} e^{{\rm i} k_1 
(S^+_{\x_1} - \theta S^+_{\x_1}) \beta^{1/2} /2 n^{1/2}}  \dots 
\right] \phantom{\int_{R^{2nl}} } \!\!\!\!\! \!\!\!\!\!\!\!\!\! 
\right).
\end{eqnarray}
In terms of the partition function $Z(h)$, this means that
\beq
|Z(h_{\rm max})|^2 \leq Z(h^{(1)}) Z(h^{(2)}),
\eeq
where $h^{(1)}$ and $h^{(2)})$ are obtained from $h_{\rm max}$ by 
reflection across $M$, in the following way:
\beq
h^{(1)}_{\x\y} = \left\{ \begin{array}{ll} h_{\x\y} & {\rm if\ } \x, 
\y\in \Lambda_L \\ h_{\vartheta\x\vartheta\y} & {\rm if\ } \x, \y\in 
\Lambda_R \\ 0 & {\rm if\ } \langle \x \y\rangle \in M \end{array} 
\right.
\eeq
and $h^{(2)}$ is given by the same expression, interchanging $L$ and $R$.
Therefore also $h^{(1)}$ and $h^{(2)}$ must be maximizers of $Z(h)$. 
However, one of them will contain strictly more zeros than $h_{\rm 
max}$, since $h_{\rm max}$ does not vanish identically for bonds 
crossing $M$. This contradicts our assumption that $h_{\rm max}$ 
contains the maximal number of zeros among all maximizers of $Z(h)$. 
Hence $h_{\rm max}\equiv 0$ identically. This completes the proof of 
(\ref{gauss}).

The next step is to transfer the upper bound on the Duhamel two point 
function (\ref{infb}) into an upper bound on the
thermal expectation value. This involves convexity arguments and
estimations of double commutators like in Section~3 in \cite{DLS}.
For this purpose, we have to evaluate the double commutators
\begin{equation}
   [\widetilde S^1_\vecp,[H,\widetilde S^1_{-\vecp}]]+
[\widetilde S^2_\vecp,[H,\widetilde S^2_{-\vecp}]]=-\frac 2 {|\Lambda|}
       \Big(H - \half \lambda|\Lambda| + 2\sum_{\langle
\vecx\vecy\rangle} S^3_\vecx
S^3_\vecy \cos \vecp\cdot(\vecx-\vecy)\Big).
\end{equation}
Let $C_\vecp$ denote the expectation value of this last expression,
$$
C_\vecp= \langle [\widetilde S^1_\vecp,[H,\widetilde
S^1_{-\vecp}]]+[\widetilde S^2_\vecp,[H,\widetilde S^2_{-\vecp}]]
\rangle\geq 0.
$$
The positivity of $C_\vecp$ can be seen from an
eigenfunction-expansion of the trace.  {F}rom \cite[Corollary~3.2 and
Theorem~3.2]{DLS} and (\ref{infb}) we infer that
\begin{equation}\label{dlsb}
\langle \widetilde S_\vecp^1 \widetilde S_{-\vecp}^1 + \widetilde S_\vecp^2
\widetilde S_{-\vecp}^2\rangle\leq \frac 12 \sqrt {
\frac  {C_\vecp}{E_\vecp}} \coth \sqrt{\beta^2 C_\vecp E_\vecp /4}.
\end{equation}
Using $\coth x \leq 1+1/x$ and Schwarz's inequality, we obtain for the
sum over all $\vecp\neq \0$,
\begin{equation}
     \sum_{\vecp\neq \0}\langle \widetilde S_\vecp^1 \widetilde S_{-\vecp}^1 +
      \widetilde S_\vecp^2 \widetilde S_{-\vecp}^2\rangle \leq \frac 1{\beta}
      \sum_{\vecp\neq \0} \frac 1{E_\vecp} + \frac 12 \Big(
\sum_{\vecp\neq \0} \frac
      1{E_\vecp} \Big)^{1/2} \Big( \sum_{\vecp\neq \0} C_\vecp 
\Big)^{1/2}. \label{sump}
\end{equation}
We have $\sum_{\vecp\in\Lambda^*} C_\vecp = -2 \langle H
\rangle+\lambda|\Lambda|$, which can
be bounded from above using the following lower bound on the Hamiltonian:
\beq\label{lowH}
H\geq -\frac {|\Lambda|}4\left
[d(d+1)+4\lambda^2\right]^{1/2}+\half\lambda|\Lambda|.
\eeq
This inequality follows from the fact that the lowest eigenvalue of
\begin{equation}\label{sumt}
-\frac 12 S_\vecx^1 \sum_{i=1}^{2d} S_{\vecy_i}^1-\frac 12 S_\vecx^2
\sum_{i=1}^{2d} S_{\vecy_i}^2+ \lambda S_\vecx^3
\end{equation}
is given by $-\mbox{$\frac 14$}[d(d+1)+4\lambda^2]^{1/2}$. This can 
be shown exactly in the same way as
\cite[Theorem~C.1]{DLS}. Since the Hamiltonian $H$ can be written as 
a sum of terms like
(\ref{sumt}), with $\vecy_i$ the nearest neighbors of $\vecx$, we get from
this fact the lower bound (\ref{lowH}).

With the aid of the sum rule
$$
\sum_{\vecp\in \Lambda^*}\langle \widetilde S_\vecp^1 \widetilde S_{-\vecp}^1 +
\widetilde S_\vecp^2
\widetilde S_{-\vecp}^2\rangle=\frac {|\Lambda|}2
$$
(which follows from $(S^1)^2=(S^2)^2=1/4$), we obtain from 
(\ref{sump}) and (\ref{lowH}) the following lower
bound in the thermodynamic limit:
\begin{eqnarray}\nonumber
      &&\lim_{\Lambda\to \infty} \frac 1{|\Lambda|} \langle \widetilde S_\0^1
      \widetilde S_{\0}^1 + \widetilde S_\0^2 \widetilde S_{\0}^2\rangle\\
      &&\geq \frac 12 -\frac12 \left(\half
        \left[d(d+1)+4\lambda^2\right]^{1/2} c_d \right)^{1/2} - \frac
      1{\beta} c_d, \label{frrom}
\end{eqnarray}
with $c_d$ given by
\beq
c_d=\frac 1{(2\pi)^{d}} \int_{ [-\pi,\pi]^d} d\vecp\frac 1{E_\vecp} .
\eeq
This is our final result. Note that $c_d$ is finite for $d\geq 3$.
Hence the right side of (\ref{frrom}) is positive, for large enough 
$\beta$, as long as
$$
\lambda^2 < \frac 1{c_d^2}-\frac {d(d+1)}4.
$$
In $d=3$, $c_3\approx 0.505$ \cite{DLS}, and hence this condition is 
fulfilled  for
$\lambda \lesssim 0.960$.  In \cite{DLS} it was also shown that $dc_d$
is monotone decreasing in $d$, which implies a similar result for all
$d>3$.

The connection with BEC is as follows.
Since $H$ is real, also $\gamma(\vecx,\vecy)$ is real and we have
$$
\gamma(\vecx,\vecy)= \langle S_\vecx^+ S_\vecy^ - \rangle = \langle
S_\vecx^1 S_\vecy^1+S_\vecx^2 S_\vecy^2
\rangle.
$$
Hence, if
$\varphi_0=|\Lambda|^{-1/2}$ denotes the constant function,
$$
\langle \varphi_0 |\gamma| \varphi_0\rangle =  \langle \widetilde S_\0^1
      \widetilde S_{\0}^1 + \widetilde S_\0^2 \widetilde S_{\0}^2\rangle,
$$
and thus the bound (\ref{frrom}) implies that the largest eigenvalue 
of $\gamma(\x,\y)$ is bounded from below by the right side of 
(\ref{frrom}). In addition one can show that the infrared
bounds imply that there is at most {\it one} large eigenvalue (of the 
order $|\Lambda|$), and that the corresponding eigenvector (the \lq 
condensate wave function\rq) is strictly constant in the 
thermodynamic limit \cite{ALSSY}.
The constancy of the condensate wave function is surprising and is
not expected to hold for densities different from $\half$, where
particle-hole symmetry is absent.
In contrast to the condensate wave function the particle density shows the
staggering of the periodic potential \cite[Thm.~3]{ALSSY}. It also 
contrasts with the situation
for zero interparticle interaction, as discussed at the end of this paper.

\bigskip

In the BEC phase there is {\it no gap} for adding particles beyond half
filling (in the thermodynamic limit): The ground state energy, 
$E_{k}$, for $\half|\Lambda|+k$
particles satisfies
\beq\label{enfin}
0\leq E_{k}-E_{0}\leq\frac{(\rm const.)}{|\Lambda|}
\eeq
(with a constant that depends on $k$ but not on $|\Lambda|$.) The 
proof of~(\ref{enfin})
is by a variational calculation, with a trial state of the form 
$(\widetilde S^+_\0)^k |0\rangle$, where $|0\rangle$ denotes the 
absolute ground state, i.e., the ground state for half filling. (This 
is the unique ground state of the Hamiltonian, as can be shown using 
reflection positivity. See Appendix~A in \cite{ALSSY}.)
Also, in the thermodynamic limit, the energy per site for a given 
density, $e(\varrho)$, satisfies
\beq\label{enth}
e(\varrho)-e(\half)\leq \const (\varrho - \half)^2.
\eeq
Thus there is no cusp at $\varrho=1/2$. To show this, one takes a 
trial state of the form
\beq
|\psi_\y\rangle= e^{{\rm i}\varepsilon \sum_\x S_\x^2} (S_\y^1+\half)|0\rangle.
\eeq
The motivation is the following: we take
the ground state and first project onto a given direction of $S^1$ on
some site $\vecy$. If there is long-range order, this should imply
that essentially all the spins point in this direction now. Then we
rotate slightly around the $S^2$-axis. The particle number should then
go up by $\eps|\Lambda|$, but the energy only by $\eps^2|\Lambda|$.
We refer to \cite[Sect.~IV]{ALSSY} for the details.

The absence of a gap in the case of BEC is not surprising, since a 
gap is characteristic for a Mott insulator state. We show the 
occurrence of a gap, for large enough $\lambda$, in the next section.

\section{Absence of BEC and Mott Insulator Phase}

The main results of this section are the following:
If either
\begin{itemize}
\item  $\lambda \geq 0$ and $T> d/(2 \ln 2)$, or
\item  $T\geq 0$ and $\lambda \geq 0$  such that $\lambda + |e(\lambda)|> d$,
with $e(\lambda)=$ ground state energy per site,
\end{itemize}
then there is exponential decay of correlations:
\beq
\gamma(\x,\y)\leq{\rm (const.)}\exp(-\kappa|\x-\y|)
\eeq
with $\kappa>0$. Moreover, for $T=0$, the ground state energy in a 
sector of fixed particle number $N=\half |\Lambda|+k$, denoted by 
$E_k$, satisfies
\beq
E_k + E_{-k} -2 E_0 \geq (\lambda + |e(\lambda)|-
d)|k| .
\eeq
I.e, for large enough $\lambda$ the chemical potential  has a jump at 
half filling.



The derivation of these two properties is based on a path 
integral representation of the equilibrium state at temperature $T$, 
and of the 
ground state which is obtained in the limit $T\to 
\infty$.  
density matrix.  
The analysis starts from the observation 
that the density operator $e^{-\beta H}$ has non-negative 
matrix 
elements in the basis in which $\{ S_{\x}^3\}$ are diagonal, 
i.e. of 
states with specified particle occupation numbers.   
It is 
convenient to focus on the dynamics 
of the `quasi-particles' which 
are defined so that the presence 
of one at a site $\x$ signifies a 
deviation there from the occupation state which minimizes the 
potential-energy.  
Since the Hamiltonian is $H=H_0+\lambda W$,
with $H_0$ the hopping term in (\ref{hamspin}) and $W$ the staggered
field, we define the quasi-particle number operators $n_\x$ as: 
\beq
n_\x \ = \  \half+(-1)^\x S_{\x}^3=\begin{cases}
  a^\dagger_{\x}a_{\x} & \hbox{for $\x$ even} \\
    1-a^\dagger_{\x}a_{\x} & \hbox{for $\x$ odd}\end{cases} \,  .
\eeq
Thus $n_{\x}=1$ means presence of a particle if $\x$ is on the
A sublattice (potential maximum) and absence if $\x$ is on the
B sublattice (potential minimum). 

The collection of the joint eigenstates of the occupation numbers, $\{
|\{n_\x\}\rangle \}$, provides a convenient basis for the Hilbert
space.  The functional integral representation of $\langle \{n_\x\}
|\, e^{-\beta (H_0+\lambda W)} \, |\{n_\x\}\rangle$ involves an integral
over configurations of quasi-particle loops in a {\em space $\times$
time} for which the (imaginary) `time' corresponds to a variable with
period $\beta$.  The fact that the integral is over a positive measure
facilitates the applicability of statistical-mechanics intuition and
tools.  One finds that the quasi-particles are suppressed by the
potential energy, but favored by the entropy, which enters this
picture due to the presence of the hopping term in $H$.  At large
$\lambda$, the potential suppression causes localization: long
`quasi-particle' loops are rare, and the amplitude for long paths
decays exponentially in the distance, both for path which may occur
spontaneously and for paths whose presence is forced through the
insertion of sources, i.e., particle creation and annihilation
operators.  Localization is also caused by high temperature, since the
requirement of periodicity implies that at any site which participates
in a loop there should be be at least two jumps during the short
`time' interval $[0,\beta)$ and the amplitude for even a single jump
is small, of order $\beta$.

The path integral described above is obtained through the 
Dyson 
expansion
\begin{equation}\label{dyson}
    e^{t(A+B)}=e^{tA}\sum_{m\geq 0}\int_{0\leq t_{1}\leq
    t_{2}\leq \cdots \leq t_m\leq t}B(t_{m})\cdots B(t_{1})dt_{1}\cdots dt_{m}
\end{equation}
for any matrices $A$ and $B$ and $t>0$, with $B(t)=e^{-tA}Be^{tA}$. 
(The $m=0$ term in the sum is interpreted here as $1$.) 

In 
evaluating the matrix elements of $e^{-\beta H} \ = \ 
e^{-\beta 
(H_0+\lambda W)}$,  in the basis $\{  |\{n_\x\}\rangle \}$,  
we note 
that $W$ is diagonal and  
$\langle \{n_\x\}| H_0 |\{n'_\x\}\rangle$ 
is non-zero only if the configurations  $\{n_\x\}$ and $\{n'_\x\}$ 
differ at exactly 
one nearest neighbor pair of sites where the 
change corresponds 
to either a creation of a pair of quasi-particles 
or the 
annihilation of such a pair.   
I.e., the matrix elements are 
zero unless $n_\x=n_\x'$ for all $\x$ 
except for a nearest neighbor 
pair $\langle \x\y\rangle$, 
where $n_\x=n_\y$, $n'_\x=n'_\y$, and 
$n_\x+n'_\x=1$. 
In this case, the matrix element equals 
$-1/2$.

Introducing intermediate states, the partition function can 
thus be
written as follows:
\begin{eqnarray}\nonumber
\Tr\, e^{-\beta H} \!\!\! &=& \!\!\! \sum_{m=0}^\infty \int_{0\leq t_{1}\leq
    t_{2}\leq \cdots\leq t_m \leq  \beta} 
\sum_{|\{n^{(i)}_\x\}\rangle, \, 1\leq i \leq m}
\\ \nonumber && \times \exp\left( -\lambda  \sum_{i=1}^{m} 
(t_{i}-t_{i-1}) \sum_\x n_\x^{(i)} \right)dt_{1}\cdots dt_{m}
\\ \nonumber && \times (-1)^m \langle \{n^{(1)}_\x\} | H_0 | 
\{n^{(m)}_\x\}\rangle  \langle \{n^{(m)}_\x\} |H_0| 
\{n^{(m-1)}_\x\}\rangle \\ &&\quad \times \langle 
\{n^{(m-1)}_\x\}|H_0|   \{n^{(m-2)}_\x\}\rangle \cdots  \langle 
\{n^{(2)}_\x\}|H_0| |\{n^{(1)}_\x\}\rangle   \label{part}
\end{eqnarray}
with the interpretation $t_0=t_m-\beta$. Note that the factor in the
last two lines of (\ref{part}) equals $(1/2)^m$ if adjacent elements
in the sequence of configurations $\{n^{(i)}_\x\}$ differ by exactly
one quasi-particle pair, otherwise it is zero.

\begin{figure}[htf]
\center
\includegraphics[width=11cm, height=6cm]{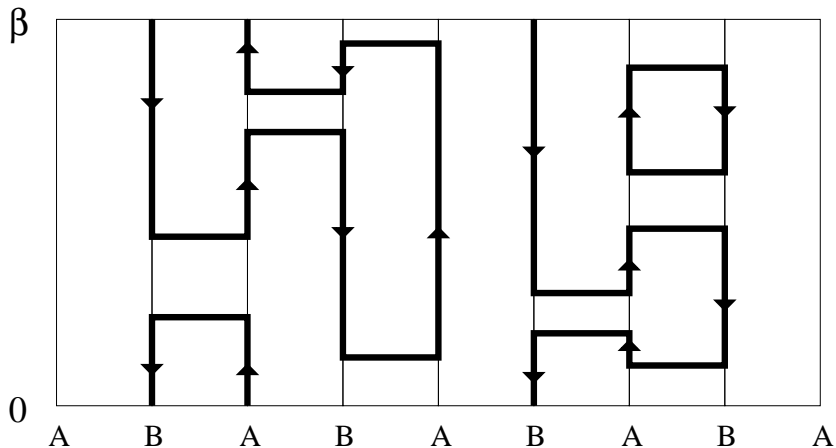}
\caption{Loop gas describing paths of quasi-particles for particle
     number $N=|\Lambda|/2-1$. A line on an A site means presence of a
     particle, while on a B site it means absence. The horizontal rungs
     correspond to hopping of a particle.}
\label{loopfig}
\end{figure}

Expansions of this type are explained more fully in \cite{AN}. A
compact way of writing (\ref{part}) is:
\beq \label{pathi}
\Tr\, e^{-\beta H}= \int
v(d\omega)e^{-\lambda|\omega|}.
\eeq
Here the \lq path\rq\ $\omega$
stands for a set of disjoint oriented loops in the \lq space-time\rq\
$\Lambda \times [0,\beta]$, with periodic boundary conditions in \lq
time\rq. Each $\omega$ is parametrized by a number of jumps, $m$,
jumping times $0\leq t_1\leq t_2\leq \dots \leq t_m \leq \beta$, and a
sequence of configurations $\{n^{(i)}_\x\}$, which is determined by
the initial configuration $\{n^{(1)}_\x\}$ plus a sequence of \lq
rungs\rq\ connecting nearest neighbor sites, depicting the creation 
or annihilation of a pair of neighboring quasi-particles (see 
Fig.~\ref{loopfig}).  
As in Feynmann's picture of QED, 
it is convenient to 
regard such an event as a jump of the 
quasi-particle, at which its 
time-orientation is also reversed. 
The length of $\omega$,
denoted by $|\omega|$, is the sum of the vertical lengths of the loops.
The measure $v(d\omega)$ is determined
by (\ref{part}); namely, for a given sequence of configurations
$\{n^{(i)}_\x\}$, $1\leq i\leq m$, the integration takes places over
the times of the jumps, with a measure $(1/2)^m dt_1\cdots dt_m$. 

One may note that the measure $v(d\omega)$ corresponds to a
Poisson process of random configurations of oriented `rungs', linking 
neighboring sites at random times, and signifying either the 
creation 
or the annihilation of a pair of quasiparticles.  
The 
matrix element 
$\langle \{n_\x\}| e^{-\beta H} |\{n'_\x\}\rangle$ 
gets no contribution from 
rung configurations that are inconsistent, 
either internally or with 
the boundary conditions corresponding to 
the specified state vectors. 
A consistent configuration yields a 
family of non-overlapping loops 
which describe the motion of the 
quasi-particles in in the 
`space-time' $\Lambda\times [0,\beta)$. 
Each such configuration 
contributes with weight 
$e^{-\lambda|\omega|}$ to the above matrix 
element (another positive 
factor was absorbed in 
the measure $v(d\omega)$). 
One may note 
that long paths are suppressed in the integral  (\ref{gamma1}) at a 
rate which increases with $\lambda$.

Likewise, for $\x\neq \y$, we can write
\beq
\Tr \,a^\dagger_{\x}a_{\y}e^{-\beta
H}=\int_{ {\mathcal A}^{(\x,\y)}} v(d\omega)e^{-\lambda|\omega|},
\eeq
where ${\mathcal A}^{(\x,\y)}$ denotes the set of all loops that, 
besides disjoint closed loops, contain one curve which avoids all the 
loops and connects $\x$ and $\y$ at time zero.
The one-particle density matrix can thus be written
\beq\label{gamma1}
\gamma(\x,\y)=\frac{\int_{{\mathcal A}^{(\x,\y)}} 
v(d\omega)e^{-\lambda|\omega|}}{\int v(d\omega)e^{-\lambda|\omega|}}.
\eeq

For an upper bound, we can drop the condition in the numerator 
that the loops and the curve from $\x$ to $\y$ do not intersect. The 
resulting measure space is simply a Cartesian product of the measure 
space appearing in the denominator and the space of all curves, 
$\zeta$, connecting $\x$ and $\y$, both at time 0. Denoting the 
latter by ${\mathcal B}(\x,\y)$, we thus get the upper bound
\beq\label{gamma2}
\gamma(\x,\y)\leq \int_{{\mathcal B}(\x,\y)}
v(d\zeta)e^{-\lambda|\zeta|} .
\eeq

The integral over paths is convergent if either $\lambda$ or 
$T$ is 
small enough, and away from the convergence threshold the 
resulting 
amplitude decays exponentially.  A natural random 
walk estimate, see 
\cite[Lemma~4]{ALSSY}, leads
to the claimed exponential bound provided
\beq
d \left(1-e^{-\beta\lambda}\right) < \lambda.
\eeq
This includes, in particular, the cases $T>d$ for any $\lambda$, and 
$\lambda > d$ for any $T$. 




Exponential decay actually holds for the larger range of parameters where
\beq\label{condim}
d \left(1-e^{-\beta(\lambda-f)}\right) < \lambda - f,
\eeq
where $f=f(\beta,\lambda)=-(\beta|\Lambda|)^{-1} \ln \Tr e^{-\beta 
H}$ is the free energy per site.
Note that $f<0$. This condition can be obtained by a more elaborate 
estimate than the one used in obtaining
(\ref{gamma2}) from (\ref{gamma1}), as shown in 
\cite[Lemma~3]{ALSSY}. The argument there uses reflection positivity 
of the measure $v(d\omega)$. Using simple bounds on $f$ one can then 
obtain  from (\ref{condim}) the conditions stated in the beginning of 
this section.

\bigskip

The proof of the energy gap is based on an estimate for the ratio
$\frac{{\rm Tr}\,{\mathcal P}_{k}e^{-\beta H}}{{\rm Tr}\,{\mathcal 
P}_{0}e^{-\beta
     H}}$ where ${\mathcal P}_{k}$ projects onto states in Fock space
with particle number $N=\half|\Lambda|+k$, expressing numerator and
denominator in terms of path integrals.
The integral for the numerator is over configurations $\omega$ with a
non-trivial winding number $k$. Each such configuration includes a
collection of \lq non-con\-tract\-ible\rq\ loops with total length at least
$\beta |k|$. An estimate of the relative weight of such loops yields 
the bound
\beq
\frac{{\rm Tr}\,{\mathcal P}_{k}e^{-\beta H}}{{\rm Tr}\,{\mathcal 
P}_{0}e^{-\beta
H}}\leq
{\rm (\const)}(|\Lambda|/|k|)^{|k|}\left(e^{1-{\rm
(const.)}\beta}\right)^{|k|}
\eeq
which gives for $\beta\to \infty$
\beq
E_k-E_{0}\geq {\rm (\const)}|k|
\eeq
independently of $|\Lambda|$. We refer to \cite{ALSSY} for details.

\section{The Non-Interacting Gas}\label{sectfree}

The interparticle interaction is essential for the existence of a Mott
insulator phase for large $\lambda$. In case of absence of the
hard-core interaction, there is BEC for any density and any $\lambda$
at low enough temperature (for $d\geq 3$). To see this, we have to
calculate the spectrum of the one-particle Hamiltonian $-\half\Delta +
V(\vecx)$, where $\Delta$ denotes the discrete Laplacian and
$V(\vecx)=\lambda (-1)^\vecx$. The spectrum can be easily obtained by
noting that $V$ anticommutes with the off-diagonal part of the
Laplacian, i.e., $\{ V, \Delta+2d\} = 0$. Hence \beq
\left(-\half\Delta - d + V(\vecx) \right)^2 = \left(-\half \Delta -
   d\right)^2 + \lambda^2, \eeq so the spectrum is given by \beq d\pm
\sqrt{\left(\mbox{$\sum_i$} \cos p_i\right)^2 +\lambda^2}, \eeq where
$\vecp\in\Lambda^*$. In particular, $E(\vecp)-E(0)\sim \half d
(d^2+\lambda^2)^{-1/2} |\vecp|^2$ for small $|\vecp|$, and hence there
is BEC for low enough temperature. Note that the condensate wave
function is of course {\it not} constant in this case, but rather
given by the eigenfunction corresponding to the lowest eigenvalue of
$-\half\Delta+\lambda(-1)^\vecx$.

\section{Conclusion}

In this paper a lattice model is studied, which is similar to the
usual Bose-Hubbard model and which describes the transition between
Bose-Einstein condensation and a Mott insulator state as the strength
$\lambda$ of an optical lattice potential is increased.  While the
model is not soluble in the usual sense, it is possible to prove
rigorously all the essential features that are observed
experimentally.  These include the existence of BEC for small
$\lambda$ and its suppression for large $\lambda$, which is a
localization phenomenon depending heavily on the fact that the Bose
particles interact with each other.  The Mott insulator regime is
characterized by a gap in the chemical potential, which does not exist
in the BEC phase and for which the interaction is also essential. It
is possible to derive bounds on the critical $\lambda$ as a function
of temperature.

\end{document}